\documentstyle[12pt]{article}

\newcommand{\be}{\begin{equation}}
\newcommand{\ee}{\end{equation}}
\newcommand{\bea}{\begin{eqnarray}}
\newcommand{\eea}{\end{eqnarray}}
\newcommand{\NP}[1]{Nucl. Phys.\ {\bf #1}\ }
\newcommand{\PL}[1]{Phys. Lett.\ {\bf #1}\ }

\newcommand{\PR}[1]{Phys. Rev.\ {\bf #1}\ }
\newcommand{\PRL}[1]{Phys. Rev. Lett.\ {\bf #1}\ }

\newcommand{\MPL}[1]{Mod. Phys. Lett.\ {\bf #1}\ }

\def\lsim{\;\raise0.3ex\hbox{$<$\kern-0.75em\raise-1.1ex\hbox{$\sim$}}\;}
\def\gsim{\;\raise0.3ex\hbox{$>$\kern-0.75em\raise-1.1ex\hbox{$\sim$}}\;}
\newcommand{\eqref}[1]{(\ref{#1})}
\newcommand{\CR}{\nonumber \\}

\newcommand{\eg}{{\it e.g. }}
\begin{document}
{\begin{flushright}
{\bf \bf UCL-IPT-96-12}
\end{flushright}}
\begin{flushleft}
\vspace*{40mm}
{ \bf
ELECTROWEAK BARYOGENESIS AND THE MINIMAL  \\
 SUPERSYMMETRIC STANDARD MODEL
\footnote{Work supported in part by the EEC Science Project SC1-CT91-0729.}\footnote{to appear in the proceedings of the NATO ASI on Masses of Fundamental Particles-Cargese (August 5-17,1996)}}\\
\vspace{15mm}
\hspace{1in}{ D. Del\'epine\footnote{Research assistant of the National
Fund for the Scientific Research}}\\
\hspace{1in}{ Institut de Physique Th\'eorique}\\
\hspace{1in}{ Universit\'e catholique de Louvain}\\
\hspace{1in}{     B-1348 Louvain-la-Neuve, Belgium}
\end{flushleft}
\vspace{15mm}
{ \bf ABSTRACT}\\
\par In principle, the baryon asymmetry of the Universe can be generated
at the electroweak phase transition but the experimental lower limit on the
Higgs mass seems to rule out a Standard Model scenario. However, it has
been shown recently that in the Minimal Supersymmetric Standard Model, the
electroweak phase transition can be a strong enough first order one for
baryogenesis if the mass of one top squark is close to or smaller than the
top mass.
\\
\\
{\bf INTRODUCTION}\\
\par Why the Universe is dominated by matter and not by antimatter is one
of the most intriguing characteristic of the Nature\cite{review}.
Astrophysical observations imply in fact a very small value for the Baryon
Asymmetry of the Universe (BAU) :
\be
\frac{n_{B}-n_{\bar B}}{n_{\gamma}}\simeq 4-7 \ 10^{-10}
\ee
with $n_B$,$n_{\bar B}$ and $n_{\gamma}$ are respectively the densities of
baryons, antibaryons and photons.
\par In 1967, Sakharov \cite{SA} was the first to point out that the BAU
could be explained in terms of high energy physics. He showed that,in order
to generate it, a particle theory has to satisfy 3 conditions:\\
- Baryon number cannot be conserved.\\
- C and CP(C for Conjuguaison Charge and P for parity) have to be
violated. Otherwise, the rate of reactions with particles and the rate of
reactions with antiparticles should be the same.\\
- Departure from thermal equilibrium is needed. At some stage of its
history, the Universe had to be out of thermal equilibrium. During these
phases, the state of Universe was non-stationary and some macroscopic
variables as the baryonic charge was time-dependent.
\par This letter is divided in 2 parts. In the first,  properties of the
Standard Model are discussed.In the second, it will be shown that in a
simplified scenario described by the Minimal Supersymmetric Standard Model
(MSSM) the baryogenesis constraint can be satisfied. The allowed range of
the parameters of the model is consistent with the present experimental
bounds \cite{DD}.
\\
\\
{ \bf THE STANDARD MODEL}\\
\par In the electroweak Standard Model (SM), the Sakharov's conditions are
fullfilled.The SM has 2 sources of CP-violation. The first is the
CP-violation coming from the phase ($\delta_{CKM}$) in the
Cabibbo-Kobayashi-Maskawa quark mixing matrix \cite{CKM} with 3
generations of quarks. A basis-invariant measure of this is given by the
Jarlskog's determinant \cite{JK}:
\be
\Delta_{CP}= det[ M_uM_u^{\dagger},M_dM_d^{\dagger}]
\ee
where $M_u$ and $M_d$ are respectively the $3\times 3$ up and down quark
mass matrices.
\par Some recent attempts to calculate the BAU using the CKM CP-violating
phase led to the conclusion that the $\delta_{CKM}$ is not efficient enough
to produce the right order of magnitude \cite{Huet}.
\par The second source of CP-violation in the SM is the
$\theta_{strong}$-angle. The most general QCD lagrangian  contains the
following 4-dimensional term:
\be
{\cal{L}}_{\theta}=\theta_{QCD}\frac{g_s^2 N_f}{32 \pi^2}G^{\mu \nu}_a
\tilde G_{\mu \nu}^a
\ee
with $N_f$ is the flavor number,$g_s$ the QCD gauge coupling, $G^{\mu
\nu}_a $ is the gluon field strength tensor and $\tilde G_{\mu \nu}^a $,
its dual. This term is T-violating (T for time reversal) and its related to
axial $U(1)$ anomaly.\\
 Indeed, at the classical level, the QCD lagrangian for $N_f$ massless
quarks is invariant under the chiral $U{(N_f)_L}\times U{(N_f)_R}$
symmetry. But at the quantum level, the flavor-singlet axial current
$J^{\mu 5} $($\bar q \gamma^{\mu}\gamma^5q$) is not conserved.
\be
\partial_{\mu}J^{\mu 5}=\frac{g_s^2 N_f}{16 \pi^2}G^{\mu \nu}_a \tilde
G_{\mu \nu}^a
\ee
\par In the procedure to diagonalize the arbitrary mass matrices of the
standard electroweak model, a chiral redefinition of the right-handed
fields is necessary.
\be
q_R \rightarrow e^{-i \frac{\theta }{N_f} }q_R
\ee
with $\theta=argdetM_uM_d$.
\par The anomalous effect of this chiral transformation is to induce the
following modification to the lagrangian:
\be
\delta{\cal{L}}_{\theta}=\theta \frac{g_s^2 N_f}{32 \pi^2}G^{\mu \nu}_a
\tilde G_{\mu \nu}^a
\ee
\par Therefore,the physical $\theta$ is the sum of both contributions: one
from QCD (Eq.[3]) and the other from the mass matrices (Eq.[6]).
\par The strongest constraint on the physical $\theta$ is coming from the
neutron electric dipole momentum \cite{Ven}:
\be
\theta_{physical} < 10^{-10}
\ee
\par It is interesting to note that this limit is of the same order of
magnitude than the BAU.\\
\par The out of equilibrium condition will be satisfied if the Electroweak
Phase transition (EWPT) is a first order one.It means that the vacuum of
the symmetric phase is metastable.The Phase Transition (PT) proceeds by
nucleation. In such a case, 3 temperatures can be defined: one when  both
vacua are degenerated ($T_1$), a second when the PT occurs ($T_c$) and the
last one is when the potential is flat at the origin($T_0$). These
temperatures follow this hierarchy:
\be
T_0<T_c<T_1
\ee
\par In the SM, the first order phase transition is induced by the weak
gauge bosons. It is a consequence of the cubic term in the finite
temperature bosonic effective potential\cite{Dolan}.
\be
V_{bosons}(m,T)=-\frac{T^4 \pi^2 }{90}+\frac{m^2 T^2}{24}-\frac{ m^3 T
}{12 \pi }+...
\ee
\be
V_{fermions}(m,T)=- {\frac{7}{180}} \pi^2 T^4+ \frac{m^2
T^2}{12}+\frac{m^4}{16 \pi^2 } ln \frac{m^2}{T^2}+...
\ee
with m is the field-dependent mass of the bosons or the fermions.
\par As we can see, the first order character of the EWPT in the SM is
proportional to the weak gauge coupling.The phase transition is not
expected to be strong enough for baryogenesis as we shall see later.
\\
\par The last Sakharov's condition is the non-conservation of the baryonic
charge. In the SM, only the B-L current is conserved (B and L are
respectively the Baryonic and the Leptonic currents) but the divergence of
the B+L current is given by the electroweak anomaly induced by the chiral
structure of the weak gauge symmetry.
\be
\partial_{\mu}J^{\mu }_{B+L}=\frac{g_w^2 N_g}{16 \pi^2}W^{\mu \nu}_a
\tilde W_{\mu \nu}^a
\ee
with $N_g$ is the generation number,$g_w$ the weak gauge coupling, $W^{\mu
\nu}_a $ is the weak field strength tensor and $\tilde W_{\mu \nu}^a $, its
dual.
\par But at zero temperature, the rate of the anomalous B-violating
reactions ($\Gamma_B$) is strongly suppressed by an exponential factor
\cite{hooft}:
\be
\Gamma_{B} \propto e^{\frac{-1}{\alpha_w}} \approx 10^{-100}
\ee
\par This suppression can be avoided at high temperature. In that case,
$\Gamma_B$ is proportionnal to a Boltzman factor and the B-violating
transition is induced by an unstable solution of the equation of motion
called "Sphaleron" \cite{Mac}.
\be
\Gamma_B \propto e^{\frac{-M_{sph}}{T}}
\ee
where  $M_{sph} = 4 \pi v(T)/g_w  B(\lambda/g^2)$ is the sphaleron mass,
$B$ is a constant which in the standard model ranges between $1.5 \le B \le
2.7$ for $0 \le \lambda/g^2 < \infty$ and $v(T)$ is the Higgs expectation
value at the temperature T.
\par To avoid a wash-out of the BAU by the sphalerons after the phase
transition, the B-violating processes have to be out of equilibrium. A way
to impose this property is to ask that $\Gamma_B$ has to be smaller than
the expansion rate of the Universe ($\Gamma_H$):
\be
\Gamma_B < \Gamma_H
\ee
\par Using Eq.[13] and the expression of the $M_{sph}$, the last condition
can be written as follows:
\be
\frac{v(T_c)}{T_c} \geq 1
\ee
\par In the SM, this baryogenesis constraint implies an upper bound on the
Higgs mass\cite{shapo}:
\be
m_H \leq 60 GeV
\ee
 This value is experimentally ruled out\cite{data}.
\par In conclusion of this first part, we have to mention that the SM
effectively fills the 3 Sakharov's conditions but it fails on 2 main
points. First, even if the CP-violating processes producing the BAU are not
well known and understood , the SM CP-violation seems to be too small.
Secondly, at the Electroweak Phase Transition, the jump in the Higgs
expectation value is too weak. So, in order to explain the BAU, we need to
go beyond  the SM.
\\
\\
{ \bf THE MINIMAL SUPERSYMMETRIC STANDARD MODEL (MSSM)}\\
\par A simple extension of the SM is the MSSM which not only predicts a
light Higgs boson but also contains new CP-violating phases. In the MSSM,
there are 2 complex scalar doublets. But we shall assume that at
low-energy, only one neutral scalar remains light while all the other Higgs
bosons and supersymmetric partners of the SM particles have a mass of the
order of the global supersymmetry breaking scale. The tree level scalar
potential for the real component h of the lightest Higgs boson  reads
\be
V_{\rm tree}=\frac{1}{2}\mu^2h^2+\frac{1}{32}\tilde{g}^2\cos^2 2\beta \ h^4\
\ee
where $h=h_1\cos \beta +h_2\sin \beta$, $\tilde{g}^2=(g^2_y+g^2_w)$ and
$g_{y,w}$ are the $U(1)$ and $SU(2)$ gauge couplings respectively.
\par In order to simplify our discussion, we shall assume that the stop
masses are given by the following relations:
\be
m^2_{\tilde{t}_{L,R}} = m^2_{L,R}+ m^2\
\ee
with $ m \equiv \frac{g_t h}{ \sqrt{2}}$. We have neglected the D-term
contribution to the stop masses as well as the left-right mixing effects.
The top and the stop loops are the dominant contributions to the effective
1-loop potential. For the Higgs mass,one obtains
\be
m^2_h =  m^2_Z\cos^2 2\beta +\frac{3}{4\pi^2}\;\frac{m^4_t}{v_0^2}\ln \left[
\left(1+\frac{m^2_L}{m^2_t}\right) \left(1+\frac{m^2_R}{m^2_t}\right)
\right]  \
\ee
where $m_Z=\tilde{g}v_0 / 2$ is the $Z$-boson mass and
$m_t=g_tv_0/\sqrt{2}$ is the  top quark mass, $v_0=246\ {\rm GeV}$.
As we can see from Eq.[9], the strength of the phase transition can be
enhanced by the right- or left-handed stop field contribution if $m_L$ or
$m_R$ is close to zero.  A scenario with $m_R \ll m_L$ is naturally
implemented even if at GUT scale $m_R=m_L$ (universality of the soft
masses)\cite{carena}.This effect is due to the 3:2:1 hierarchy in the
renormalisation group equations for the Higgs scalar $h_2$, right-handed
and left-handed squared masses respectively\cite{nilles}. Finally, assuming
$m_R \ll m_L$  and $m_t \ll m_L$ and keeping only the relevant terms in the
effective potential, we obtain
\be
V(h,T)=M^2(T) h^2-\delta(T)h^3 - a(T) (h^2+b^2)^{3/2} + \lambda(T) h^4\  ,
\ee
where
\bea
M^2(T)&=& -\frac{1}{4}m^2_Z\cos^2 2\beta
-\frac{3}{16\pi^2}\;\frac{m^2_t}{v_0^2}\left\{ m_t^2 \ln
\left[\left(1+\frac{m^2_L}{m^2_t}\right) \left(1+\frac{m^2_R}{m^2_t}\right)
\right] \right. \CR
           &+& \left. m_L^2 \ln \left(1+\frac{m_t^2}{m_L^2}\right) + m_R^2
\ln (m_t^2+m_R^2) + \frac{1}{2} m_R^2\right\} \CR
    &+& \frac{m_t^2}{2 v_0^2} T^2 + \frac{3}{16 \pi^2} \frac{m_t^2}{v_0^2}
m_R^2 (2\ln T + C_B)\ ,
\eea
\be
\delta(T)=\frac{2 m_W^3+m_Z^3}{6\pi v_0^3} T \ ,
\ee
\be
a(T)=\frac{m_t^3 T}{2\pi v_0^3}\ , \;\; b=\frac{m_R v_0}{m_t}\ ,
\ee
\be
\lambda(T)=\frac{1}{8} \frac{m_Z^2}{v_0^2} \cos^2 2\beta -
\frac{3}{16\pi^2} \frac{m_t^4}{v_0^4}\left( \ln \frac{T}{m_L} - C_F -
\frac{1}{2} C_B -\frac{3}{4}\right)\ ,
\ee
$C_B$ and $C_F$ are constants coming from the high temperature expansion
\cite{Dolan} ($C_B=5.41$ and $C_F=-2.64$).
 The form of the potential is simple and can be analytically studied as
done in \cite{DD}.As said before, the phase transition occurs at a
temperature $T_c$ between $T_0$ and $T_1$ and these relations remain valid
for the $\frac{v(T)}{T}$:
\be
\frac{v(T_1)}{T_1} \lsim \frac{v(T_c)}{T_c} \lsim \frac{v(T_0)}{T_0}\ .
\ee
\par To conclude, the $\frac{v(T)}{T}$ ratios are plotted as a function of
tan$\beta$ for $m_{\tilde {t_R}}=m_t$ and $m_L=500$ GeV in Fig.1. We can
see that from the point of view of the baryogenesis the favourite value for
tan$\beta$ is between 0.5 and 1.5. The maximum value for the ratio is
reached for tan$\beta $=1.This corresponds to the lower value for the Higgs
mass ($\approx 60 $ GeV) which is consistent with the present experimental
data on MSSM \cite{data}.
\par Under the assumption that the right-handed stop mass is close to or
smaller than the top mass, we have shown that the baryogenesis constraint
(Eq. [15] ) can be satisfied for low value of tan$\beta$ in the MSSM. In
\cite{quiros}, similar results were obtained using a numerical analysis of
the effective potential. A 2-loop numerical analysis of this potential
\cite{espinoza}and lattice calculations \cite{lattice} confirm the
enhancement of the first order phase transition in the range of the MSSM
parameters studied in this letter.
\\
\\
{\bf ACKNOWLEDGEMENT}\\
\par These results were obtained in a fruitful collaboration with J.-M.
G\'erard, R. Gonzalez Felipe and J. Weyers.
\\

\newpage
\input{epsf.sty}
\begin{figure}[t]
\leavevmode
\begin{center}
\mbox{\epsfxsize=15.cm\epsfysize=10.cm\epsffile{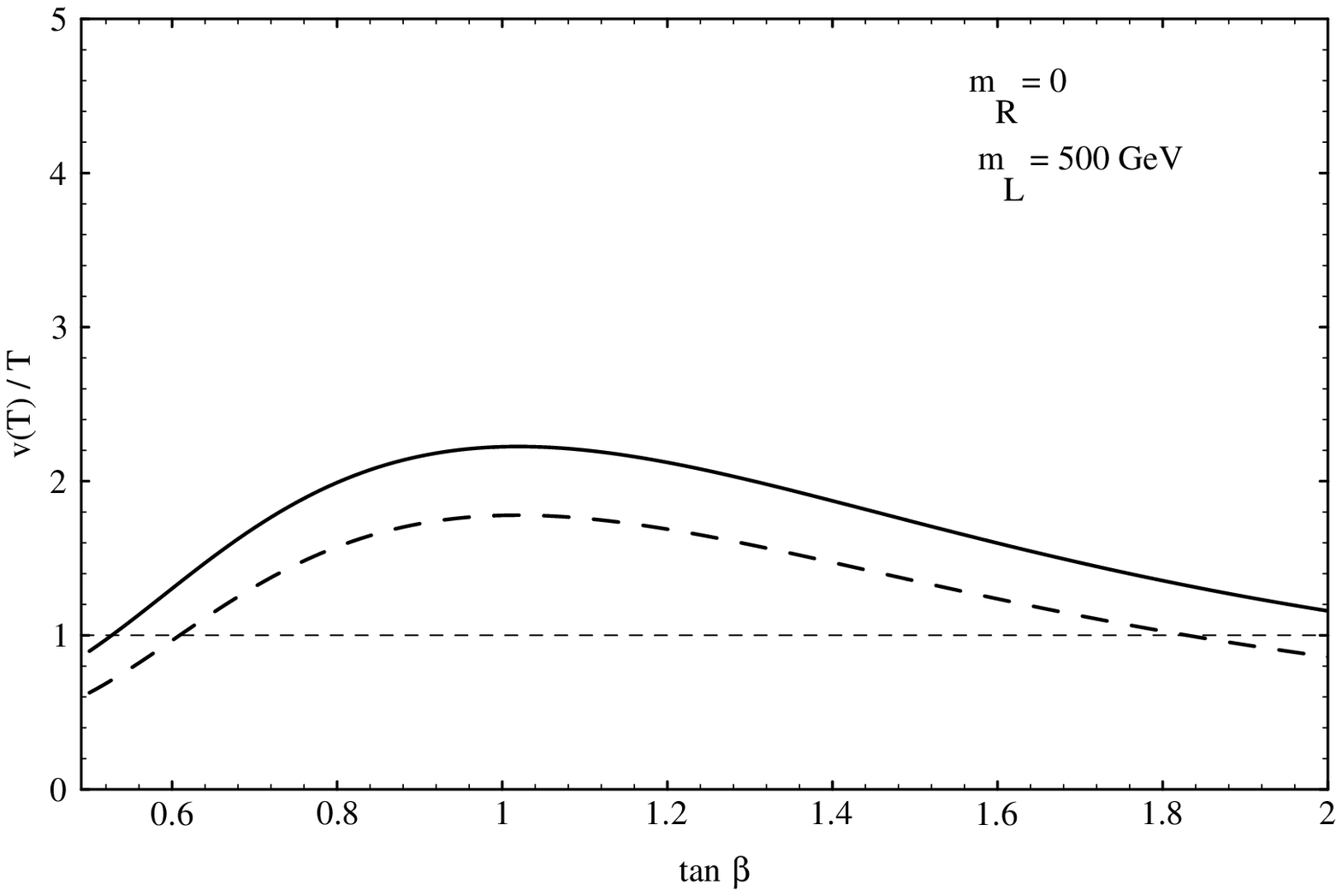}}
\end{center}
\caption{ The curves  $v(T_0)/T_0$ (solid line) and $v(T_1)/T_1$ (bold-dashed line)  as functions of $\tan \beta$ for $m_R=0$, $m_L=500$~GeV and $m_t=174$~GeV.}
\label{fig1}
\end{figure}
\end{document}